\documentclass{ws-mplb}
\usepackage{amsfonts,amssymb,amsmath}
\usepackage{graphicx}
\usepackage[super,sort,compress]{cite} 

\begin{document}

\markboth{Takao Watanabe}{Electronic phase diagram of Fe$_{1+y}$Te$_{1-x}$Se$_{x}$ revealed by magnetotransport measurements}

%
\catchline{}{}{}{}{}
%

\title{Electronic phase diagram of Fe$_{1+y}$Te$_{1-x}$Se$_{x}$ revealed by magnetotransport measurements}

\author{\footnotesize Takao Watanabe\footnote{e-mail: twatana@hirosaki-u.ac.jp}, Takumi Otsuka, Shotaro Hagisawa, Yuta Koshika, Shintaro Adachi, Tomohiro Usui, Nae Sasaki, Seya Sasaki, Shunpei Yamaguchi, Yu Uezono}

\address{Graduate School of Science and Technology, Hirosaki University, 3 Bunkyo, Hirosaki, 036-8561 Japan}

\author{Yoshiki Nakanishi, Masahito Yoshizawa}

\address{Graduate School of Engineering, Iwate University, Morioka 020-8551, Japan}

\author{Shojiro Kimura}

\address{Institute for Materials Research, Tohoku University, 2-1-1 Katahira, Aoba-ku, Sendai, 980-8577 Japan}

\maketitle

\begin{history}
\received{7 February 2020}
\accepted{20 March 2020}
\end{history}

\begin{abstract}
Among the Fe-based superconductors, Fe$_{1+y}$Te$_{1-x}$Se$_{x}$ is unique in that its crystal structure is the simplest and the electron correlation level is the strongest, and thus it is important to investigate the doping($x$)-temperature ($T$) phase diagram of this system. However, inevitably incorporated excess Fe currently prevents the establishment of the true phase diagram. We overcome the aforementioned significant problem via developing a new annealing method termed as ``Te-annealing'' wherein single crystals are annealed under Te vapor. 
Specifically, we conducted various magnetotransport measurements on Te-annealed superconducting Fe$_{1+y}$Te$_{1-x}$Se$_{x}$. We observed that crossover from the incoherent to the coherent electronic state and opening of the pseudogap occurs at high temperatures ($\approx$ 150 K for $x$ = 0.2). This is accompanied by a more substantial pseudogap and the emergence of a phase with a multi-band nature at lower temperatures (below $\approx$ 50 K for $x$ = 0.2) before superconductivity sets in. Based on the results, the third type electronic phase diagram in Fe-based high-$T_c$ superconductors is revealed.
\end{abstract}


\section{Introduction}

Knowledge of the doping($x$)-temperature ($T$) phase diagram is important in understanding the mechanism of high superconducting transition temperature (high-$T_c$) superconductivity. In Fe-based high-$T_c$ superconductors, it is widely known that two types of materials exhibit a distinct phase diagram. The first type corresponds to the ``1111" system in which antiferromagnetic (AFM) and superconducting (SC) states are separated by a first-order like transition and do not coexist.\cite{luetkens} The second type corresponds to the ``122" system. In this case, both states overlap and coexist in some $x$- and $T$-range.\cite{nandi} Among the Fe-based high-Tc superconductors, the iron chalcogenide Fe$_{1+y}$Te$_{1-x}$Se$_{x}$\cite{fang} (``11" system) is unique because its crystal structure is the simplest and the electron correlation level is the strongest.\cite{yin} Therefore, it is important to investigate the phase diagram of the aforementioned system. However, the existence of excess iron has prevented the establishment of the true phase diagram.

Recently, an O$_{2}$-annealing technique was developed to remove excess iron from the Fe$_{1+y}$Te$_{1-x}$Se$_{x}$ system. Additionally, the technique was employed to investigate its phase diagram ~\cite{tamegai, sun}. The pioneering study indicated that the AFM ordered phase and SC phase exist separately with the phase boundary at x = 0.05. Furthermore, the temperature dependence of the Hall coefficient, $R_{H}$, showed a peak at $T^{*}$. The result suggests that a phase of a multi-band nature appears below $T^{*}$. However, this anomalous behavior is not yet fully understood to date.

Conversely, several angle-resolved photoemission spectroscopy (ARPES) studies focus on this compound.~\cite{ieki,shen} The results indicate that the evolution from incoherent to coherent electronic states occurs with increases in Se concentrations. The samples with x $\le$ 0.2 corresponded to non-SC while those with x $\ge$ 0.4 corresponded to SC, thereby suggesting a close relationship between the coherent electronic state and the SC. However, an excess amount of iron may be incorporated in these ARPES samples, and thus the aforementioned studies did not focus on the effects of excess iron and doping (i.e., the value of x). Therefore, it is necessary to more systematically examine the crossover phenomenon and establish the phase diagram.

In the study, we review extant studies on this subject. First, we overcome the serious problem of excess Fe by developing a new annealing method termed as ``Te-annealing".~\cite{koshika} Subsequently, electron transport measurements are performed on the Te-annealed crystals and the as-grown crystals.~\cite{otsuka} Based on the obtained results, the third type electronic phase diagram in Fe-based high-$T_c$ superconductors in which AFM and SC states exist separately via a second-order like transition while incoherent to coherent crossover always occurs at higher temperatures than $T_c$ provided that the sample is a bulk SC, will be revealed. 

\section{Crystal growth and Te-annealing}

\begin{figure*}[th]
		\begin{center}
		\centerline{\includegraphics[width=12cm]{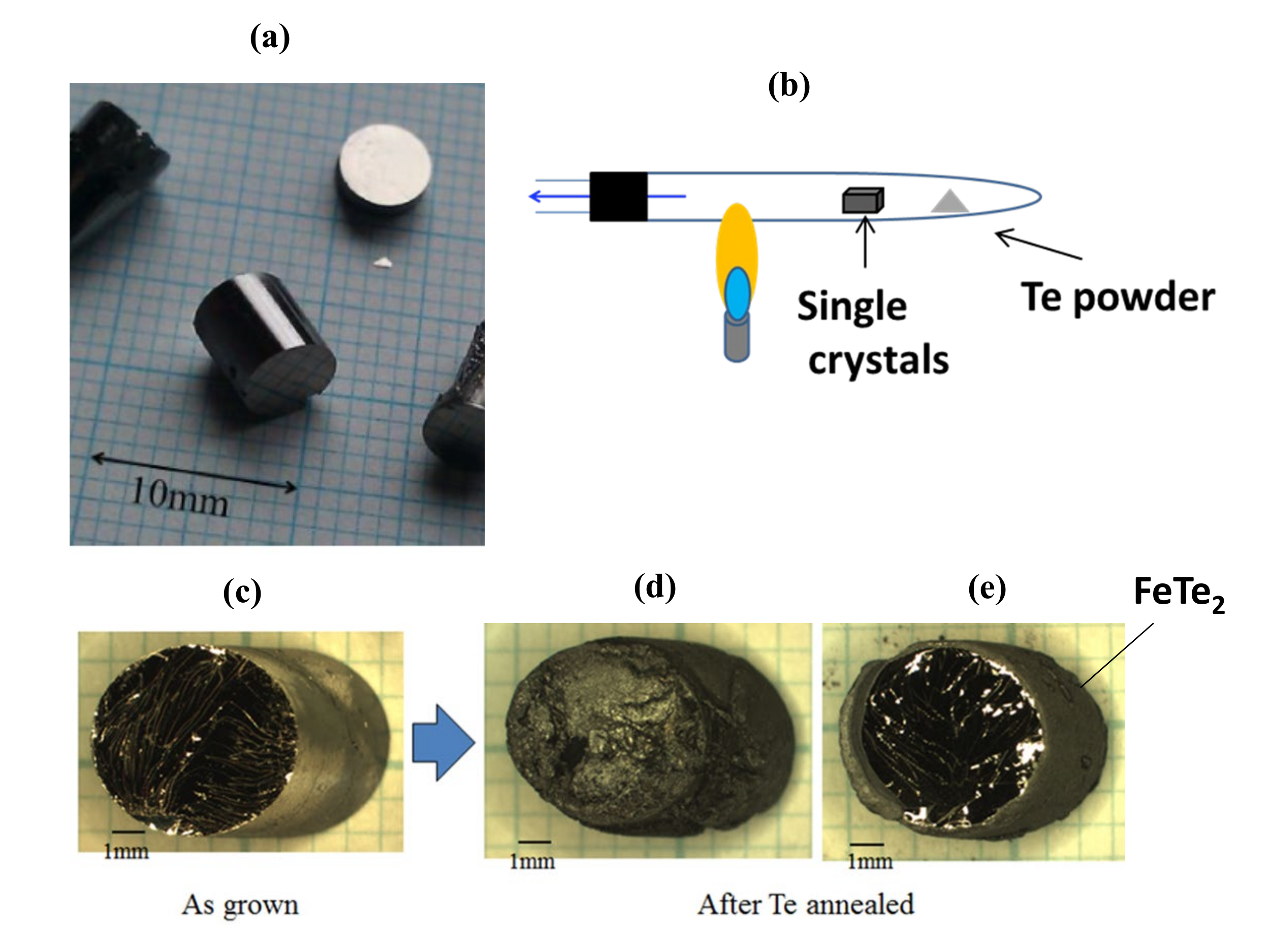}}
			\caption{\label{fig1}(Color online)  (a) Photograph of as-grown Fe$_{1.03}$Te$_{0.8}$Se$_{0.2}$ single crystals.~\cite{koshika} (b) Schematic drawing of the Te-annealing method. (c) As-grown single crystal before Te-annealing. (d) Single crystal after Te-annealing. (e) Te-annealed single crystal after cleaving away the grayish top surface portion. The grayish compound (FeTe$_2$) remains at the side .} 
		\end{center}
	\end{figure*}
	
Single crystals of Fe$_{1+y}$Te$_{1-x}$Se$_{x}$ (0$\le x \le$0.4) were grown via the Bridgman method (the slow cooling method under a temperature gradient of around 8$^\circ$C/cm by using a vertical furnace with 3-zones).~\cite{koshika} Large-sized (6 mm$\phi$ x 20 mm) and high quality single crystals are obtained as shown in Fig. \ref{fig1}(a). Subsequently, to remove excess Fe, we developed a new annealing method termed as ``Te-anneal" in which single crystals are annealed under Te vapor.~\cite{koshika} The as-grown crystals were cleaved into smaller crystals of $\approx$ 1-mm thickness, and they were loaded into a glass tube with an appropriate amount of Te powder. The glass tube was held for more than 400 h at 400 $^\circ$C. The Te-annealing method is shown schematically in Fig. \ref{fig1}(b). After Te-annealing, the surface of the as-grown crystal (Fig. \ref{fig1}(c)) became grayish (Fig. \ref{fig1}(d)). However, after cleaving away the surface portion, a shiny and thick crystal appears (Fig. \ref{fig1}(e)), and this enables us to measure out-of-plane resistivity $\rho_{c}$ and in-plane resistivity $\rho_{ab}$. Our chemical composition analysis using an electron probe microanalyzer (EPMA) indicated that the excess Fe existed in the as-grown crystal was completely removed by Te-annealing.~\cite{koshika} Furthermore, the temperature dependence of the magnetic susceptibilities $\chi$ of the Te-annealed Fe$_{1+y}$Te$_{1-x}$Se$_{x}$ single crystals exhibited sharp superconducting transitions ($\Delta$ $T_{c}$ $\le$ 1 K) with onset temperatures corresponding to 12.1, 12.8, 13.7, and 14.5 K, for $x$ = 0.15, 0.2, 0.3, and 0.4, respectively.~\cite{otsuka} The data confirm that the superconducting properties of the Te-annealed samples are very good. Additionally, it should be noted that all the samples exhibit the Meissner signal (negative value in the FC data) although the signal is very weak, and this can be attributed to the sufficient removal of excess Fe by Te-annealing. The grayish part of the surface (Fig. \ref{fig1}(d)) was observed as FeTe$_2$ by X-ray diffraction (XRD) measurements.~\cite{koshika} Therefore, we assume that during Te-annealing the excess Fe atoms escape from inside of the crystal to participate in forming FeTe$_2$ with surrounding Te atoms at the crystal surface.

\section{Electrical transport properties}
	
\subsection{In-plane and out-of-plane resistivity} 

\begin{figure*}[t]
		\begin{center}
			\includegraphics[width=130mm]{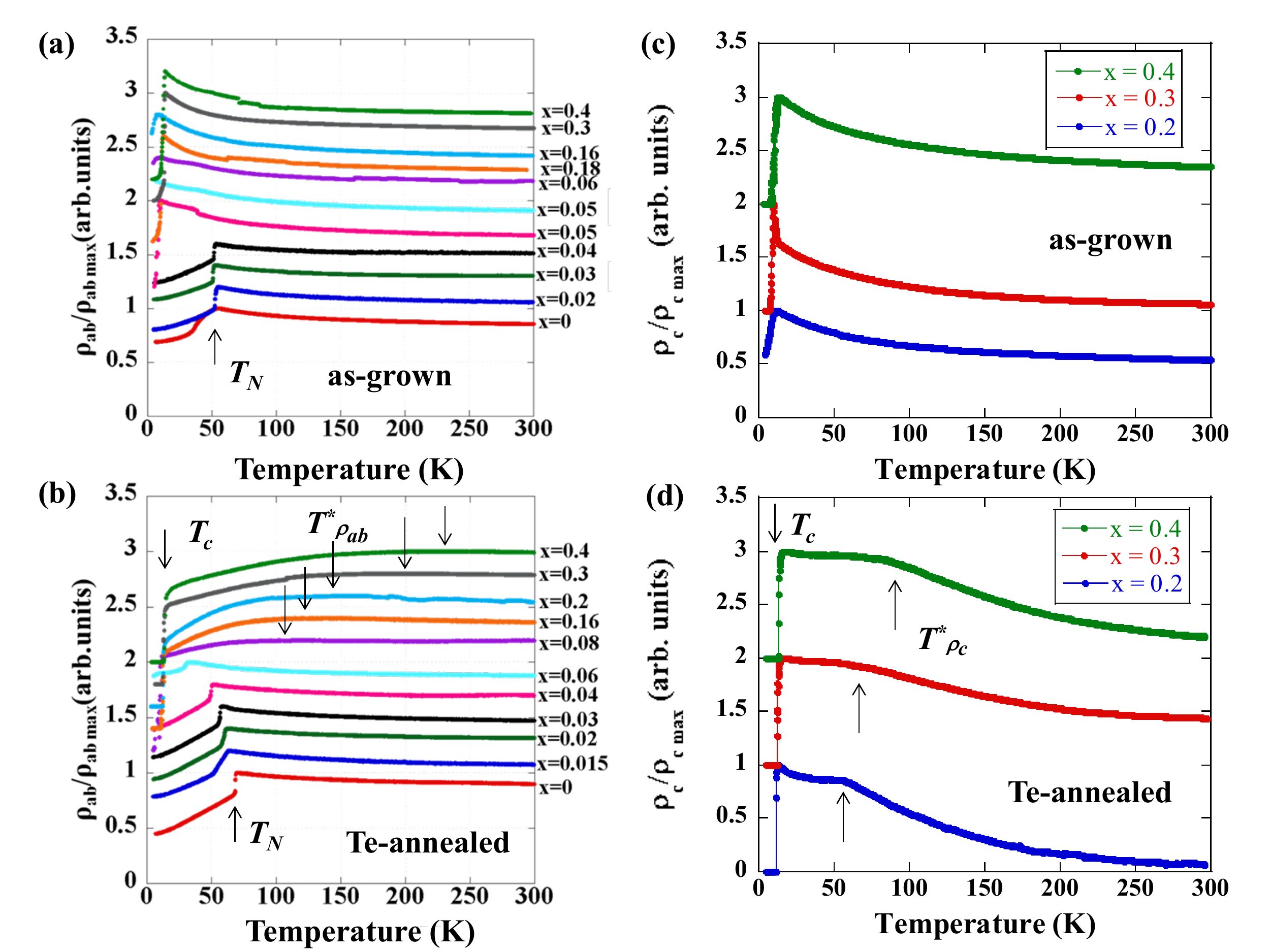}
			\caption{\label{fig2}(Color online)  Normalized in-plane resistivity $\rho_{ab}(T)$ for the (a) as-grown and (b) Te-annealed Fe$_{1+y}$Te$_{1-x}$Se$_{x}$ (0 $\le$ x $\le$ 0.4) single crystals, and normalized out-of-plane resistivity $\rho_{c}(T)$ for the (c) as-grown and (d) Te-annealed Fe$_{1+y}$Te$_{1-x}$Se$_{x}$ (0.2 $\le$ x $\le$ 0.4) single crystals.~\cite{otsuka} Resistivity data are shifted vertically for purposes of clarity. The arrows denote the superconducting, AFM transitions, and characteristic temperatures, $T^*_{\rho_{ab}}$ and $T^*_{\rho_{c}}$.}   
		\end{center}
	\end{figure*}

Figures \ref{fig2}(a) and (b) show $\rho_{ab}(T)$ for the as-grown and fully Te-annealed Fe$_{1+y}$Te$_{1-x}$Se$_{x}$ (0$\le$x$\le$0.4) single crystals, respectively.~\cite{otsuka} The as-grown crystals exhibit a resistivity drop that originated in the long-range AFM transition at $\approx$ 50 K in the doping region of 0$\le$x$\le$0.04. For $x \ge$0.05, they exhibit filamentary superconductivity~\cite{koshika}. For the annealed crystals, the AFM transition is observed at $T_{N} \approx$ 70 K for $x$ = 0, and it then decreases with increases in x to $\approx$ 30 K for x = 0.06, and bulk superconductivity appears for $x$ $\ge$ 0.08~\cite{koshika}. It should be noted that the temperature dependence of $\rho_{ab}$ for the as-grown crystals is semiconducting while that of all the Te-annealed superconducting crystals appears as a poorly resolved broad structure. Specifically, the temperature at which $\rho_{ab}$ reaches its maximum is defined as $T^*_{\rho_{ab}}$. $T^*_{\rho_{ab}}$ wherein it linearly increases from 110 K for x = 0.08 to 230 K for x = 0.4. 

Figures \ref{fig2}(c) and (d) show $\rho_{c}(T)$ for the as-grown and fully Te-annealed Fe$_{1+y}$Te$_{1-x}$Se$_{x}$ (x = 0.2, 0.3, and 0.4) single crystals, respectively. The as-grown crystals exhibit smooth semiconducting behavior before the filamentary superconducting transition. Conversely, the annealed crystals exhibit semiconducting behavior at higher temperatures similar to the as-grown crystals. However, they exhibit a typical plateau below the temperatures, $T^*_{\rho_{c}}$, before the bulk superconducting transition. Specifically, $T^*_{\rho_{c}}$ is estimated as $\approx$ 55, $\approx$ 65, and $\approx$ 90 K for x = 0.2, 0.3, and 0.4, respectively. Here, $T^*_{\rho_{c}}$ is estimated as the temperature at which the second-order derivative is minimized.

\subsection{Hall coefficient and Magnetoresistance}

\begin{figure}[t]
		\begin{center}
			\includegraphics[width=120mm]{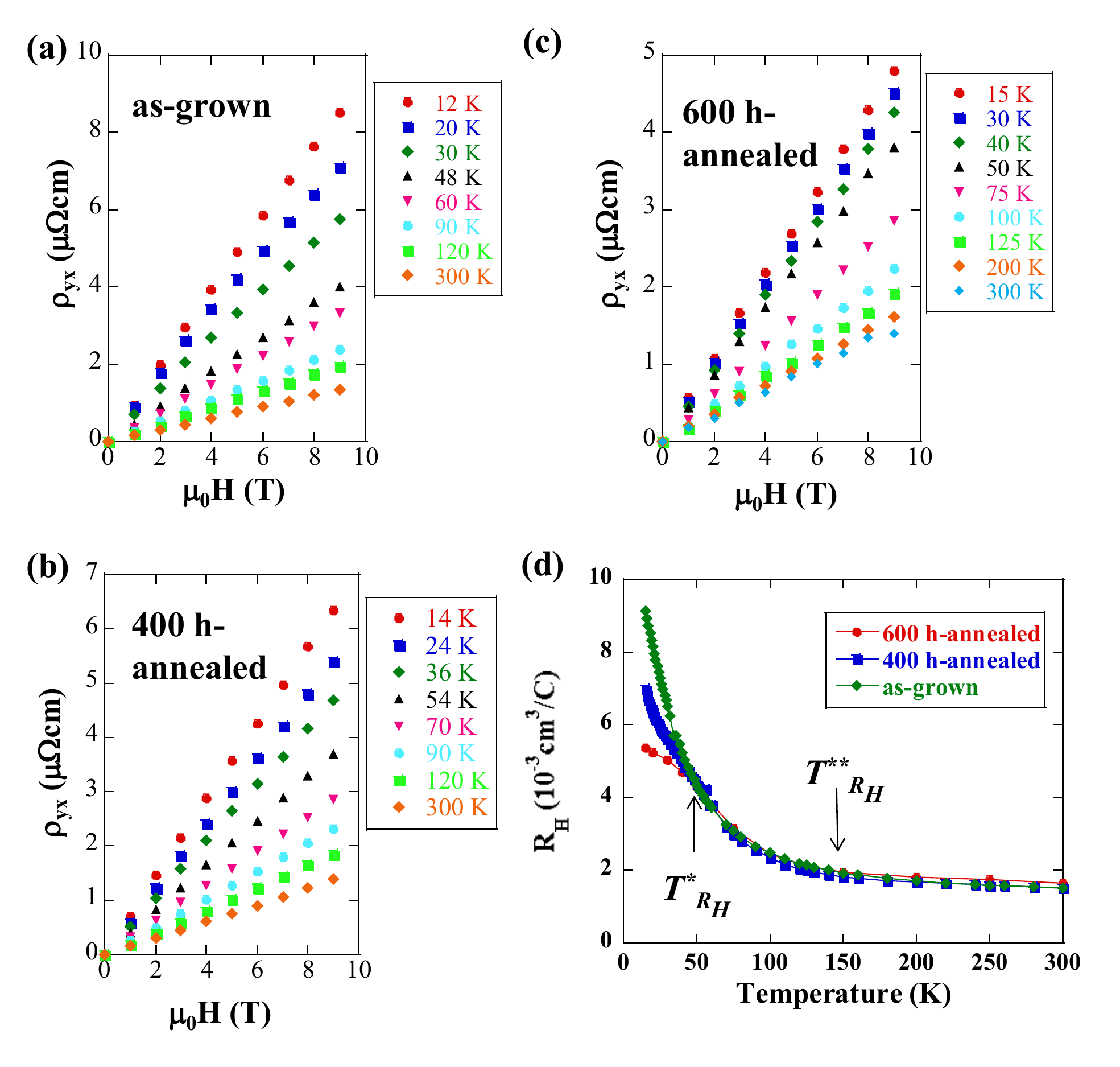}
			\caption{\label{fig3}(Color online) Hall resistivity, $\rho_{yx}$, at several temperatures for the (a) as-grown, (b) 400-h annealed and (c) 600-h annealed Fe$_{1+y}$Te$_{0.8}$Se$_{0.2}$ single crystals.~\cite{otsuka} (d) Temperature dependence of Hall coefficients, $R_{H}$, for the as-grown, 400-h annealed and 600-h annealed Fe$_{1+y}$Te$_{0.8}$Se$_{0.2}$ single crystals.~\cite{otsuka} The arrow indicates the characteristic temperature $T^*_{R_{H}}$.}
		\end{center}
	\end{figure}

In order to understand the anomalous behaviors in $\rho_{ab}(T)$ and $\rho_{c}(T)$, we measured Hall coefficients, $R_{H}$, for the Fe$_{1+y}$Te$_{0.8}$Se$_{0.2}$ single crystals. Hereafter, our discussion focuses on the measurements of x = 0.2 crystals. Figure \ref{fig3}(a)--\ref{fig3}(c) show the magnetic field dependence of the Hall resistivity, $\rho_{yx}$, at several temperatures for the as-grown, 400-h annealed, and 600-h annealed Fe$_{1+y}$Te$_{0.8}$Se$_{0.2}$ single crystals, respectively. In all cases, $\rho_{yx}$ increases linearly with the applied magnetic fields, $H$, maintaining a positive slope, $d\rho_{yx}/dH$ \textgreater 0, down to low temperatures slightly above $T_{c}$. The results indicate that the hole-type carrier dominates the electron transport. As widely-known, Fe$_{1+y}$Te$_{1-x}$Se$_{x}$ exhibits multi-bands with the hole and the electron bands near $\Gamma$ and the $M$ point, respectively.~\cite{ieki,shen} In a two-band model, $R_{H}$ is generally described as,~\cite{smith}

\begin{equation}
R_{H} = \frac{1}{e}\frac{(\mu_{h}^{2}n_{h} - \mu_{e}^{2}n_{e}) + (\mu_{h}\mu_{e})^2(\mu_{0}H)^2(n_{h} - n_{e})}{(\mu_{h}n_{h} + \mu_{e}n_{e})^{2} + (\mu_{h}\mu_{e})^2(\mu_{0}H)^2(n_{h} - n_{e})^{2}}. 
\end{equation}

where $\mu_{h}$ ($\mu_{e}$) denotes the mobility of holes (electrons) and $n_{h}$ ($n_{e}$) denotes the number of holes (electrons). Since the $\rho_{yx}$ is expressed as $\rho_{yx} = (\mu_{0}H)R_{H}$, the field dependence of $\rho_{yx}$ becomes nonlinear when $n_{h} \neq n_{e}$. Only when $n_{h} \approx n_{e}$, $\rho_{yx}$ is expected to exhibit field-linear behavior. Therefore, our observation of the field-linear behavior in $\rho_{yx}$ (Fig. \ref{fig3}(a)-\ref{fig3}(c)) indicates that Fe$_{1+y}$Te$_{0.8}$Se$_{0.2}$ can be considered as a compensated semimetal ($n_{h} \approx n_{e}$). This fact was recently confirmed from the estimation of the Fermi-surface area via ARPES.~\cite{koshiishi1} 

Subsequently, $R_{H} = \rho_{yx}/\mu_{0}H$ for the as-grown, 400-h annealed, and 600-h annealed crystals are plotted as a function of temperature and are shown in Fig. \ref{fig3}(d). For all the samples, $R_{H}$ exhibits an almost constant positive value from room temperature to $\approx$ 150 K. Below the aforementioned temperature, it gradually increases with decreases in temperature. Conversely, below 50 K, each of the samples were observed to exhibit different behavior. The result agrees with that in the previous report\cite{tamegai} and suggests that the multi-band nature manifests itself below 50 K. The temperatures at which $R_{H}$ begins to increase ($\approx$ 150 K) and ramifies ($\approx$ 50 K) are defined as $T^{**}_{R_{H}}$ and $T^*_{R_{H}}$, respectively. Notably, $T^{**}_{R_{H}}$ and $T^*_{R_{H}}$ coincide with $T^*_{\rho_{ab}}$ and $T^*_{\rho_{c}}$, respectively.

Additional insights into the anomalous $R_{H}$ were obtained via measuring the magnetoresistance (MR) for the crystals (x = 0.2). In all cases, MR is negative, and it is almost constant from room temperature to $T^{**}_{R_{H}}$ while it gradually decreases below $T^{**}_{R_{H}}$ and rapidly decreases below $T^*_{R_{H}}$.~\cite{otsuka} The behavior corresponds well with the temperature dependence of $R_{H}$. We consider~\cite{otsuka} that below $T^{**}_{R_{H}}$ ($\approx$ 150 K), because the pseudogap gradually opens, carrier density decreases and increases in $R_{H}$. Additionally, MR is negative because the pseudogap is destroyed by the magnetic fields and causes the recovery of the carrier density. Specifically, a pseudogap opening below 150 K was directly observed by ARPES in the electron band around the M point for a Te-annealed sample with $x$ = 0.2.~\cite{koshiishi1}

\subsection{Two-band analysis}

\begin{figure}[t]
		\begin{center}
			\includegraphics[width=100mm]{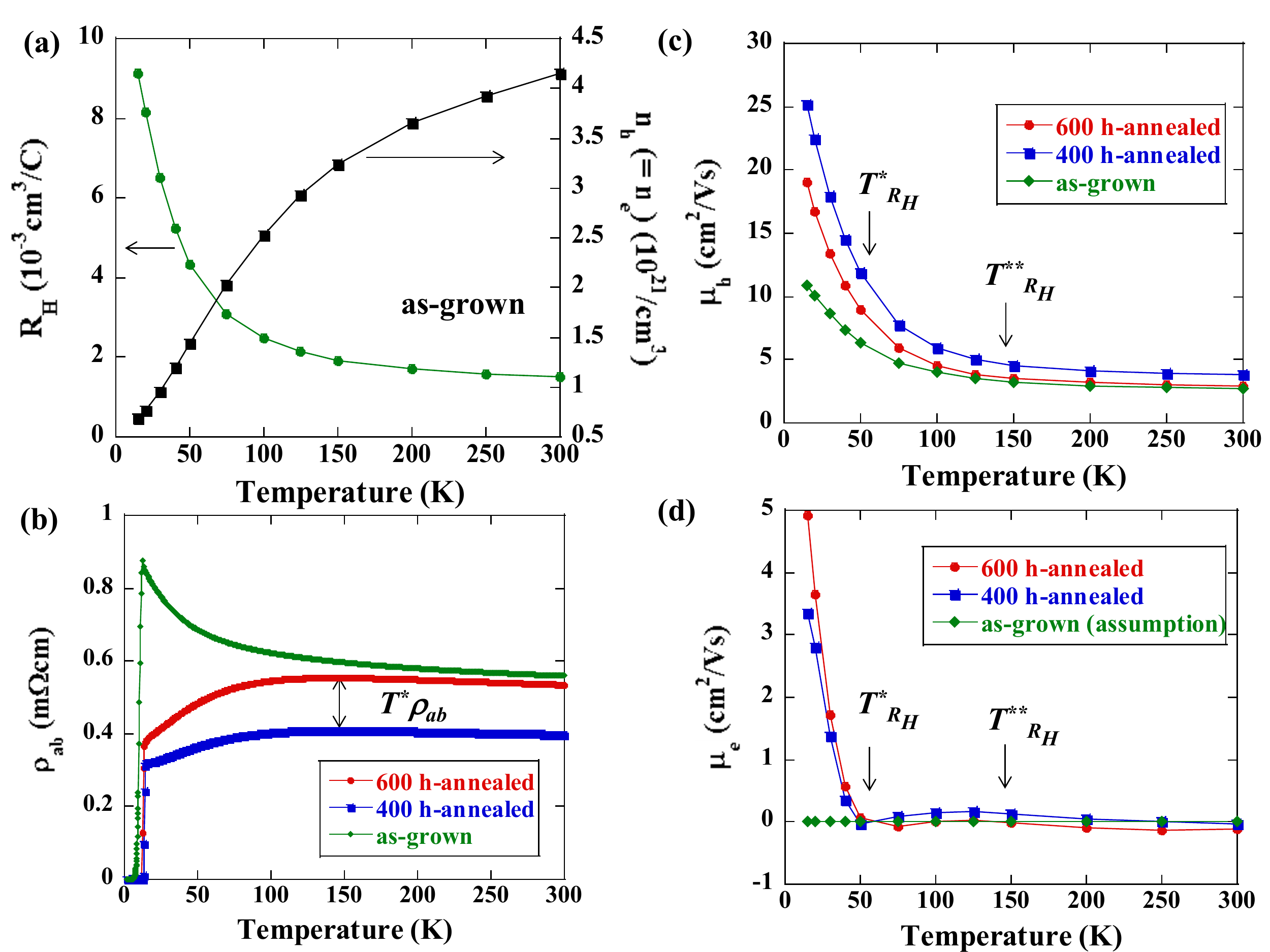}
			\caption{\label{fig4}(Color online) (a) Temperature dependence of Hall coefficient, $R_{H}$, and the hole number, $n$, for the as-grown Fe$_{1+y}$Te$_{0.8}$Se$_{0.2}$ single crystal.~\cite{otsuka} (b) In-plane resistivities, $\rho_{ab}(T)$, for the as-grown, 400-h annealed, and 600-h annealed Fe$_{1+y}$Te$_{0.8}$Se$_{0.2}$ single crystals.~\cite{otsuka} (c) Temperature dependence of hole mobilities, $\mu_{h}$, for the as-grown, 400-h annealed, and 600-h annealed Fe$_{1+y}$Te$_{0.8}$Se$_{0.2}$ single crystals.~\cite{otsuka} (d) Temperature dependence of electron mobilities, $\mu_{e}$, for the as-grown, 400-h annealed, and 600-h annealed Fe$_{1+y}$Te$_{0.8}$Se$_{0.2}$ single crystals. The arrows denote the characteristic temperatures, $T^*_{\rho_{ab}}$, $T^{**}_{R_{H}}$, and $T^*_{R_{H}}$.~\cite{otsuka}}
		\end{center}
	\end{figure}

Given that the origin of the anomalous transport properties may be the pseudogap effect, we examined what occurs at $T^{**}_{R_{H}}$ or $T^{*}_{R_{H}}$. Thus, we adopt a simple two-band model with the following three assumptions. First, we assume that FeTe$_{1-x}$Se$_{x}$ is a compensated semimetal with an equal number $n$ of electrons and holes albeit permitting $n$ to be temperature dependent. Second, we assume that $\mu_{e}$ of the as-grown sample corresponds to zero. Third, we assume that $n$ is common for all the samples. Subsequently, the in-plane resistivity and Hall coefficient are described as, $\rho_{ab} = \frac{1}{ne(\mu_{e}+\mu_{h})}$ and $R_{H} = \frac{\mu_{h}-\mu_{e}}{ne(\mu_{e}+\mu_{h})}$, respectively. Given the second assumption, this simple two-band model effectively results in a one-band model for the as-grown sample. Subsequently, $n$ can be estimated using the observed $R_{H}$ as $n = \frac{1}{eR_{H}}$ (Fig. \ref{fig4}(a)). By using the obtained $n$, $\mu_{h}$ and $\mu_{e}$ are numerically calculated using $\rho_{ab}$ (Fig. \ref{fig4}(b)) and $R_{H}$ (Fig. \ref{fig3}(d)) with the aforementioned two-band model. The results are plotted in Fig. \ref{fig4}(c) and (d), respectively. Evidently, the values of $\mu_{h}$ of the annealed samples increase significantly faster than those of the as-grown sample below 150 K ($\approx$ $T^{**}_{R_{H}}$). The result implies that hole bands of the annealed samples become coherent below $T^{**}_{R_{H}}$. Additionally, below 50 K ($\approx$ $T^{*}_{R_{H}}$), the values of $\mu_{e}$ of the annealed samples appear. The result implies that electron bands of the annealed samples become coherent below $T^{*}_{R_{H}}$. This type of incoherent to coherent crossover below $T^{**}_{R_{H}}$ and the appearance of electron pockets below $T^{*}_{R_{H}}$ is observed by ARPES measurements.~\cite{koshiishi1}

\section{Electronic phase diagram}	

\begin{figure}[t]
		\begin{center}
			\includegraphics[width=85mm]{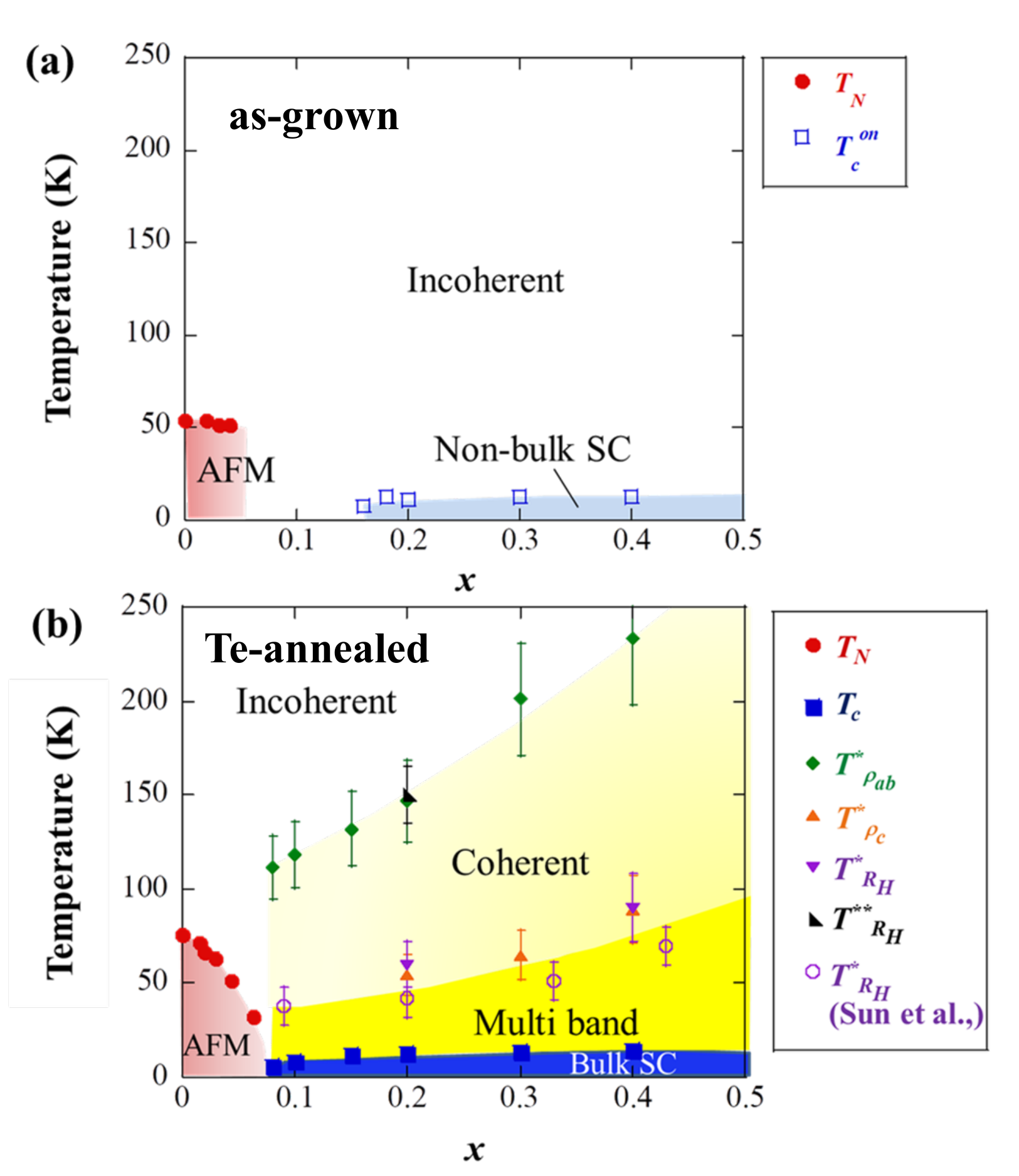}
			\caption{\label{fig5}(Color online) (a) Characteristic temperatures vs. Se concentration $x$ for as-grown Fe$_{1+y}$Te$_{1-x}$Se$_{x}$ single crystals. (b) Characteristic temperatures vs. Se concentration $x$ for Te-annealed Fe$_{1+y}$Te$_{1-x}$Se$_{x}$ single crystals.~\cite{otsuka}}
		\end{center}
	\end{figure}
	
The characteristic temperatures $T^*_{\rho_{ab}}$, $T^*_{\rho_{c}}$, $T^{**}_{R_{H}}$, $T^*_{R_{H}}$, $T_{N}$, $T_{c}^{on}$ (onset temperature for $T_{c}$), and $T_{c}$ are plotted as a function of the Se concentration $x$ for as-grown and Te-annealed samples in Fig. \ref{fig5} (a) and (b), respectively. In as-grown samples, $T_{N}$ is observed in 0$\le$x$\le$0.04, whereas bulk SC did not occur for $x \ge$0.05. In the doping range 0.05$\le$x$\le$0.16, samples can be inhomogeneously mixed with AFM and filamentary SC phases. It should be noted that the electronic states are always incoherent in the as-grown samples. Conversely, in the Te-annealed samples, the long-range AFM ordered state exists in the range of 0 $\le$ $x$ $\le$ 0.06 and the superconducting state emerges at $x$ $\ge$ 0.08. The AFM and superconducting states do not coexist agrees with the previous report~\cite{tamegai, sun}, and this implies that the double-stripe type AFM order at the parent compound Fe$_{1+y}$Te~\cite{bao} competes with superconductivity. The result appears different from the results of previous studies~\cite{tamegai, sun} because the transition is considerably gradual (second-order-like). Therefore, the obtained phase diagram of Te-annealed Fe$_{1+y}$Te$_{1-x}$Se$_{x}$ single crystals differs from that of the ``1111" system. \cite{luetkens} 

As noted above in Sec. 3.2, $T^{**}_{R_{H}}$ coincides with $T^*_{\rho_{ab}}$. This implies that incoherent to coherent crossover shown by $T^*_{\rho_{ab}}$ triggers the opening of the pseudogap which manifests in $T^{**}_{R_{H}}$. This is potentially because the Fermi surfaces become well defined when the electronic states become coherent, and this causes the pseudogap to open through inter-band nesting. Furthermore, $T^*_{R_{H}}$ and $T^*_{\rho_{c}}$ coincide. The participation of electron carriers in charge transport causes $T^*_{R_{H}}$, and thus this is also expected to be the cause for anomalous $T^*_{\rho_{c}}$. In the annealed samples, $\rho_{ab}$ is metallic while $\rho_{c}$ is semiconducting in the temperature range between $T^*_{\rho_{ab}}$ and $T^*_{\rho_{c}}$. This is very similar to the behavior of the cuprates.~\cite{usu} Conversely, $\rho_{c}$ exhibits a plateau (Fig. \ref{fig2} (d)) when the electron band appears below $T^*_{\rho_{c}}$. This is because the metallic out-of-plane conductivity originated in the electron band is added to the semiconducting conductivity of the hole bands.

In addition to the aforementioned difference in the relation between AFM and SC orders, a comparison of the phase diagram of as-grown (non-SC) samples (Fig. \ref{fig5} (a)) with that of annealed samples (Fig. \ref{fig5} (b)) clearly exhibits a new feature wherein SC samples always experience incoherent to coherent crossover at higher temperatures before the SC sets in. It should be noted that it is not possible to recognize $T^*_{\rho_{ab}}$ in the non-SC samples with low Se concentrations (x $\le$ 0.06) even when they are annealed. Thus, the $x$-$T$ phase diagram of the annealed samples should correspond to the third type in Fe-based high-$T_c$ superconductors, which is inherent to the strongly correlated Fe$_{1+y}$Te$_{1-x}$Se$_{x}$ system (``11" system).

\section{Summary}

An effective annealing method (Te-annealing) is developed to overcome the problem of excess Fe in Fe$_{1+y}$Te$_{1-x}$Se$_{x}$ system. Subsequently, the $x$-$T$ phase diagram of Fe$_{1+y}$Te$_{1-x}$Se$_{x}$ (0 $\le x \le$ 0.4) was examined via performing various transport measurements on samples with and without Te-annealing. In the superconducting samples (Te-annealed samples with 0.08 $\le x \le$ 0.4), the incoherent to coherent crossover transition, which is accompanied by the opening of the pseudogap, was observed to occur in two steps. First, only the in-plane state becomes coherent with the pseudogap opening gradually at higher temperatures below $T^{**}_{R_{H}}$ (and $T^*_{\rho_{ab}}$ $\approx$ 150 K for $x$ = 0.2). Second, the in-plane and out-of-plane states were coherent with the pseudogap opening significantly at lower temperatures below $T^*_{R_{H}}$ (and $T^*_{\rho_{c}}$ $\approx$ 50 K for $x$ = 0.2). Based on the obtained phase diagram, we propose that the incoherent to coherent crossover is a prerequisite for the occurrence of SC in this system.

\section*{Acknowledgments}

We thank K. Koshiishi, A. Fujimori, H. Kontani, S. Onari, T. Tamegai, and Y. Koike for their helpful discussions. This work was supported by Hirosaki University Grant for Distinguished Researchers FY2017-2018. The magnetotransport measurements
		were mostly performed using PPMS at Iwate University. Some of the measurements were performed at the High Field Laboratory for Superconducting Materials, Institute for Materials Research, Tohoku University (Project No. 14H0007).

\end{document}